\documentstyle[prl,twocolumn,aps,epsfig]{revtex}
\def\prn#1{{\left(#1\right)}}
\def\cbrk#1{{\left\{#1\right\}}}
\def\sbrk#1{{\left[#1\right]}}
\def\fig_width{3.375 in} % width of single column figure in PR
\draft
\begin{document}
\title{Search for exchange-antisymmetric two-photon states}
%==================================
\author{D.~DeMille$^{a}$, D.~Budker$^{b}$,
N.~Derr$^{c,}$\cite{byline2}, and E.~Deveney$^{c}$\cite{byline3}}
\address{$^a$Department of Physics, Yale University, New Haven, CT 06520 \\
$^b$Department of Physics, University of California, Berkeley, CA
94720-7300\\ and Lawrence Berkeley National Laboratory, Berkeley
CA 94720 \\
$^c$Department of Physics, Amherst College, Amherst, MA
01002 }
\date{\today}
%%% ----------------------------------------------------------------------
\maketitle
%%% ----------------------------------------------------------------------
\begin{abstract}
Atomic two-photon J=0 $\leftrightarrow$J'=1 transitions are
forbidden for photons of the same energy. This selection rule is
related to the fact that photons obey Bose-Einstein statistics. We
have searched for small violations of this selection rule by
studying transitions in atomic Ba. We set a limit on the
probability $v$ that photons are in exchange-antisymmetric states:
$v<1.2\cdot10^{-7}$.
\end{abstract}
\pacs{PACS numbers: 03.65.Bz, 05.30.Jp, 42.50.Ar, 32.80.-t} Recent
experiments have explored the possibility of small violations of
the usual relationship between spin and statistics
\cite{Ramberg,Deilamian,de Angelis,HilbornYuca,Modugno}. Although
such violations are impossible within conventional quantum field
theory \cite{Streater}, there are motivations for considering
them; see e.g. Ref. \cite{Greenberg}.  If photons do not obey
Bose-Einstein (BE) statistics, there will be a non-zero
probability that two photons are in an exchange-antisymmetric
state.  Here we report the results of a search for such states
based on a selection rule \cite{Bonin,Bowers} that forbids
two-photon transitions between atomic states with $J=0$ and $J'=1$
for degenerate photons (i.e., photons of equal energy).

Consider the general amplitude for a $J=0\rightarrow{}J'=1$
two-photon transition.  This amplitude is a scalar which must be
constructed from the polarization vectors $\mathbf{e}_{1,2}$ of
the photons and the polarization vector $\mathbf{e}_{v}$
describing the $J'=1$ state (each exactly once); and may also
contain arbitrary powers of the photon propagation directions
$\hat{\mathbf{k}}_{1,2}$, and some function of the photon energies
$\hbar\Omega_{1,2}$. In the specific case of E1-E1 transitions,
the amplitude must be independent of $\hat{\mathbf{k}}_{1,2}$, and
only one form is possible:
%--------------------------------------------------------------------
\begin{equation}
  A\;\propto\;\{\prn{\mathbf{e_{1}} \mathnormal \times \mathbf{e_{2}}} \mathnormal \cdot \mathbf{e_{v}}\} \:
  \mathnormal F\prn{\Omega_{1},\Omega_{2}}, \label{Eq1}
\end{equation}
%--------------------------------------------------------------------
which requires orthogonally-polarized photons.  If photons obey BE
statistics, this amplitude must be invariant under exchange of
labels $1 \leftrightarrow 2$. Eqn. (\ref{Eq1}) satisfies this
condition only if $ \mathnormal F\prn{\Omega_{1},\Omega_{2}}$ is
odd under exchange. Therefore the amplitude must vanish in the
degenerate case, if photons behave as normally expected.  This
argument can be readily generalized beyond the E1-E1 case
\cite{Sakurai}. For the case of counterpropagating degenerate
photons (with $\mathbf{k_{2}}=-\mathbf{k_{1}}$), the Landau-Yang
theorem states that all possible amplitudes vanish \cite{Landau}.
We consider only the E1-E1 amplitude of Eqn. (\ref{Eq1}), since
higher multipolarity transitions are too weak to be observed in
the present experiment.

For ordinary photons, the atomic two-photon-resonant transition
rate is \cite{Bonin}:
%--------------------------------------------------------------------
\begin{eqnarray}
  &W_{gf}\prn{\Omega_{1},\Omega_{2}}\propto \nonumber\\
  &|{\mathbf{e_{1a}}\mathbf{e_{2b}}}\langle{}f|Q_{ab}|g\rangle|^{2}\frac{dI_{1}}{d\Omega_{1}}\frac{dI_{2}}{d\Omega_{2}}\delta\prn{\omega_{fg}-\Omega_{1}-\Omega_{2}},
\label{Eq2}
\end{eqnarray}
%--------------------------------------------------------------------
%--------------------------------------------------------------------
\begin{eqnarray}
  &Q_{ab}\prn{\Omega_{1},\Omega_{2}}=\nonumber\\
  &d_{a}\prn{\sum_{n}\frac{|n\rangle\langle{}n|}{\omega_{ng}-\Omega_{1}}}d_{b}+d_{b}\prn{\sum_{n}\frac{|n\rangle\langle{}n|}
  {\omega_{ng}-\Omega_{2}}}d_{a}.
\label{Eq3}
\end{eqnarray}
%--------------------------------------------------------------------
Here indices $g$, $f$, and $n$ indicate ground, final, and
(virtual) intermediate states of the transition;
$\frac{dI_{1(2)}}{d\Omega_{1(2)}}$ are the spectral distributions
of light intensity; $\omega_{ij}$ are frequencies of atomic
transitions; $\mathbf{d}$ is the dipole operator; and the
subscripts $a, b$ refer to Cartesian components. Consistent with
our experimental conditions, we neglect Doppler and natural widths
compared to laser spectral widths. For a $J=0\rightarrow{}J'=1$
transition, only the irreducible rank-1 component of $Q_{ab}$ can
contribute to the matrix element \cite{Bonin,Bowers}. Thus
%--------------------------------------------------------------------
\begin{eqnarray}
  &Q_{ab}\prn{\Omega_{1},\Omega_{2}}=Q_{ab}^{(1)}=\frac{1}{2}\prn{Q_{ab}-Q_{ba}}=\nonumber\\
  &\frac{1}{2}\prn{\Omega_{1}-\Omega_{2}}\sum_{n}\frac{\cbrk{d_{a}|n\rangle\langle{}n|d_{b}-d_{b}|n\rangle\langle{}n|d_{a}}} {\prn{\omega_{ng}-\Omega_{1}}\prn{\omega_{ng}-\Omega_{2}}}.
\label{Eq4}
\end{eqnarray}
%--------------------------------------------------------------------
Eqn. (\ref{Eq4}) shows explicitly that degenerate transitions are
forbidden: $Q_{ab}\prn{\Omega_{1}=\Omega_{2}}=0$.  Also, the
transition amplitude
${\mathbf{e_{1a}}\mathbf{e_{2b}}}\langle{}f|Q_{ab}|g\rangle$ has,
as expected, the form of the rotational invariant in Eqn.
(\ref{Eq1}).

We now generalize these results to allow for violation of BE
statistics. Permutation symmetry for photons is reflected in the
plus sign between the two terms in Eqn. (\ref{Eq3}). We construct
a similar "BE-Violating" two-photon operator with a minus sign
between the terms:
%--------------------------------------------------------------------
\begin{eqnarray}
&Q_{ab}^{BV}\prn{\Omega_{1},\Omega_{2}}=\nonumber\\
&\sum_{n}\frac{\omega_{ng}-\omega_{fg}/2}{\prn{\omega_{ng}-\Omega_{1}}\prn{\omega_{ng}-\Omega_{2}}}\cbrk{d_{a}|n\rangle\langle{}n|d_{b}-d_{b}|n\rangle\langle{}n|d_{a}}.
\label{Eq5}
\end{eqnarray}
%--------------------------------------------------------------------
The transition rate becomes:
%--------------------------------------------------------------------
\begin{eqnarray}
&W_{gf}\prn{\Omega_{1},\Omega_{2}}\propto \nonumber\\
&\sbrk{|{\mathbf{e_{1a}}\mathbf{e_{2b}}}\langle{}f|Q_{ab}|g\rangle|^{2}+v|{\mathbf{e_{1a}}\mathbf{e_{2b}}}\langle{}f|Q_{ab}^{BV}|g\rangle|^{2}}
\nonumber\\
&\times\frac{dI_{1}}{d\Omega_{1}}\frac{dI_{2}}{d\Omega_{2}}\delta\prn{\omega_{fg}-\Omega_{1}-\Omega_{2}},
\label{Eq6}
\end{eqnarray}
%--------------------------------------------------------------------
where $v \ll 1$ is the BE statistics violation parameter, i.e.,
$v$ is the probability to find two photons in an antisymmetric
state. Here we explicitly include the fact that the normal and
BE-violating amplitudes cannot interfere with each other
\cite{Amado}.  Eqns. (\ref{Eq4}) and (\ref{Eq6}) summarize the
central principle of our measurement:  for monochromatic light,
the degenerate $J=0\rightarrow{}J'=1$ transition rate is due
\textit{entirely} to BE statistics violation; i.e.,
$W_{gf}\prn{\Omega_{1}=\Omega_{2}}\propto v$.

Recent theoretical discussion of possible small violations of the
spin-statistics relation has centered on the "quon algebra," which
allows continuous transformation from BE to Fermi statistics
\cite{Greenberg90}. Our heuristic argument above is reproduced in
the quon formalism: if creation/annihilation operators for photons
obey the $q$-deformed commutation relations
%--------------------------------------------------------------------
\begin{equation}
a_{k}a_{l}^{+}-qa_{k}^{+}a_{l}=\delta_{kl}, \label{Eq7}
\end{equation}
%--------------------------------------------------------------------
then in Eqn. (\ref{Eq6}), $v=\prn{\frac{1-q}{2}}^{2}$
\cite{Hilborn&Greenberg}.  Degenerate $J=0\rightarrow{}J'=1$
two-photon transition are allowed only to the extent that $q$
deviates from 1.  However, it should be noted that application of
the quon formalism to photons is questionable: relativistic quon
theories exhibit nonlocal features \cite{Greenberg91}, and there
are also unresolved questions concerning issues as basic as the
interpretation of $v$ in terms of $q$\cite{Hilbornprivate}. With
these caveats, we note that the quon formalism was used previously
to set limits on the deviation of photons from BE statistics. For
instance, Fivel calculated \cite{Fivel} that the existence of
high-intensity lasers implies that $1-q\lesssim 10^{-6}$ (although
this calculation was argued to be invalid \cite{Greenberg93}).
Recently, it was shown that statistics deviations for photons and
charged particles are linked within the quon theory
\cite{Greenberg&Hilborn}, so that the stringent limits on the
deviations from Fermi statistics for electrons \cite{Ramberg} can
be used to set a limit $1-q \lesssim 10^{-25}$.

We point out that the simple argument behind our technique makes
it possible to state limits on $v$ without reference to the quon
theory.  Similar reasoning was used to show that $v \neq 0$ leads
to the decay $Z\rightarrow{}\gamma\gamma$ \cite{Ignatiev}.
However, only $v\lesssim{}1$ can be inferred, in part because $Z$
has no direct coupling to photons. The result reported here is the
first quantitative limit on the existence of
exchange-antisymmetric states for photons.

We searched for the transition
$6s^{2}~^{1}S_{0}+2\gamma\rightarrow{}5d6d~^{3}S_{1}$ in atomic Ba
(Fig. 1).  This transition has an unusually large BE-violating
amplitude (see below).  Fig. 2 shows a schematic of the apparatus.
Light from a dye laser was split into two beams with orthogonal
linear polarizations. These beams counterpropagated through a Ba
vapor cell. The laser was tuned around the required frequency for
the degenerate two-photon transition ($\lambda=549~nm$).
Transitions were detected by observing fluorescence at
$\lambda_{fl}=436~nm$, accompanying the decay
$5d6d~^{3}S_{1}\rightarrow{}6s6p~^{3}P_{2}$.  Excess signal within
a narrow tuning range of the laser wavelength would indicate a
violation of BE statistics.  The sensitivity of the experiment was
calibrated with the same detection system, using non-degenerate
photons ($\lambda_{1}^{'}=532~nm$ and $\lambda_{2}^{'}=566~nm $)
to drive the same transition. That is, we determined the ratio of
signals for the degenerate and calibration transitions:
\begin{eqnarray}
S \equiv \frac {W_{gf} \prn {\Omega_{1} = \Omega_{2} =
\omega_{gf}/2}} {W_{gf}^{'} \prn {\Omega_{1}^{'},
\Omega_{2}^{'}}},
\label{Eqextra}
\end{eqnarray}
where the primed quantities correspond to the non-degenerate
transition. The value of $S$ determines $v$: in the limit of
monochromatic light $S \propto v$. From Eqn. (\ref{Eq6}), the
proportionality (i.e., calibration) constant includes the laser
intensities and spectral widths for both transitions, and also the
ratio of two-photon operators $R \equiv
Q_{ab}^{BV}\prn{\Omega_{1}=\Omega_{2}=\frac{\omega_{fg}}{2}}/Q_{ab}\prn{\Omega_{1}^{'},\Omega_{2}^{'}}$.
We discuss the determination of these quantities below.

The central region of the vapor cell was at $T\approx{}925~K$,
corresponding to $P(Ba)\approx5\cdot{}10^{-2}~ Torr$. The cell
contained buffer gas (He, $P\approx{}0.2~Torr$). The dye laser was
pumped by a doubled, pulsed Nd:YAG laser ($532~nm$). Both lasers
had pulse length $\approx{}7~ns$. Part of the $532~nm$ light was
split off to excite the calibration transition. Fluorescence from
the cell passed through filters onto a photon-counting
photomultiplier. Counts were recorded for $25~ns$ following the
laser pulse. The laser frequencies were determined with a
wavelength meter with accuracy $\pm 5~GHz$ and reproducibility
$\pm 2.5~GHz$.

Laser intensity was determined from separate measurements of pulse
energy and beam area. Pulse energy was measured within $\pm 15\%$;
linearity was verified by comparing measurements with two types of
detectors, and by varying beam energy with calibrated neutral
density filters.  For the degenerate transition, energies
$\approx1.5~mJ/pulse$ were used in each of the counterpropagating
beams. For the nondegenerate calibration transition, the beam
powers were reduced to avoid saturation: $\sim 100~\mu{}J/pulse$
at $566~nm$ and $\sim 1 \mu{}J/pulse$ at $532~nm$. The laser beams
had diameter $\approx2.5~mm$ ($\pm 25\%$) at the vapor cell. Laser
spectral widths (averaged over many pulses) were determined with a
scanning Fabry-Perot interferometer, and also inferred from atomic
signals. To find the dye laser bandwidth, we tuned through an
allowed, degenerate two-photon transition in Ba. Using the known
dye laser bandwidth, we determined the YAG laser bandwidth by
tuning the dye laser through the calibration transition.  We
assign uncertainties of $20-30\%$ to the bandwidths to account for
the range of values obtained. Both lasers had linewidths $\approx
3~GHz$, large compared to the transition Doppler widths ($\approx
0.3~GHz$). Fig. 3 shows a typical scan through the calibration
transition. The high laser power used for the forbidden transition
can lead to complications such as AC Stark shift and broadening,
higher-order nonlinear processes, etc., which are not present at
the lower powers used for the calibration transition.  We checked
for such effects by studying the calibration transition with
varying dye laser powers of up to $5~mJ/pulse$. A correction
factor ($1.4\pm0.3$) is applied to the relative detection
efficiency for the two transitions, to account for depletion of
fluorescence at high powers (presumably due to photoionization
\cite{comment}). We saw no evidence for line shifts or distortions
even at the highest powers.

The ratio of two-photon operators $R$ is determined as follows.
Atomic transition energies \cite{Moore} $\omega_{gn}$ and
$\omega_{nf}$ and magnitudes of dipole matrix elements
\cite{Miles,Hafner} $|\langle{}n|{\mathbf{d}}|g\rangle |$ are
known for all significant intermediate states $|n\rangle$ in the
sums of Eqns. (\ref{Eq3}) and (\ref{Eq5}). We measured magnitudes
of dipole matrix elements $|\langle{}f|{\mathbf{d}}|n\rangle |$ by
determining the lifetime of, and branching ratios of all decays
from, the state $5d6d~^{3}S_{1}$. The lifetime ($25\pm15~ns$) was
measured by recording the time evolution of fluorescence.
Branching ratios were measured by observing fluorescence through a
scanning monochromator. We find that the sums over intermediate
states in Eqns. (\ref{Eq3}) and (\ref{Eq5}) are all well
approximated by a single term, corresponding to the intermediate
state $|n\rangle=6s6p~^{1}P_{1}$.  This term has small energy
denominators, and large dipole matrix elements with both the
initial and final states. (This was the reason we used this
particular transition.) In this approximation, the matrix elements
cancel in the ratio $R$, and this quantity depends only on
accurately known atomic and photonic energies. We find $R^{2}=10
\pm 2$, with the uncertainty due to the neglected terms in the
sums.

Data for the degenerate transition were taken in three separate
runs (Fig. 4).  The laser was scanned around the nominal frequency
of the degenerate transition.  The signals were fit with a
constant background plus a peak whose width was fixed by the dye
laser spectral width.  The small constant background appears to
arise from Ba-He collision-assisted transitions, but is not fully
understood.  In all three runs, there is evidence for a
statistically significant peak above the background.  The center
frequencies are consistent with the predicted position of the
degenerate transition, and with each other.  Note that these peaks
correspond to extremely weak transitions:  although the laser
intensities were much larger for the degenerate transition than
for the calibration transition
($\frac{I_{1}I_{2}}{I_{1}^{'}I_{2}^{'}} \sim 10^{4}$), the ratio
of degenerate transition to calibration transition signals is
small ($S\sim 10^{-2})$.

We believe these peaks are due to the finite bandwidth of the dye
laser.  For light from a single laser of \textit{finite} spectral
width (centered around $\Omega_{L} = \omega_{fg}/2$), the
transition probability of Eqn. (\ref{Eq6}) does not vanish for $v
= 0$, even though $Q_{ab}\prn{\Omega_{1},\Omega_{2}}=0$ for
$\Omega_{1}=\Omega_{2}$. From the known experimental parameters
and plausible models for our laser spectra, we can predict the
size of the residual signal S due to this "bandwidth effect". The
uncertainty in this prediction for each run ($\sim60\%$) was
estimated by adding in quadrature the various uncertainties in
calibration parameters. The uncertainty in the size of the fitted
peaks themselves was smaller: $25-30\%$ for each run. Averaging
over all three runs gives the result for the ratio of the observed
peak, to the predicted size of bandwidth-effect peak:
%--------------------------------------------------------------------
\begin{equation}
  \frac{S\prn{observed}}{S\prn{predicted}}=1.5\pm0.6 .
\label{Eq8}
\end{equation}
%--------------------------------------------------------------------
That is, the observed resonances are consistent with those
expected for purely bosonic photons, due to the finite bandwidth
of the dye laser.

A violation of BE statistics would appear as a resonant signal in
excess of the peak due to the bandwidth effect. (Note that from
Eqn. (\ref{Eq6}), there is no mechanism for cancellation of the
peak when $v\neq0$.) Although the observed peak is consistent with
our predictions, we find that the size of the bandwidth-effect
peak is sensitive to details of the laser spectra.  In particular,
in some models of the spectra which are implausible but not
\textit{a priori} excluded by our data, the bandwidth-effect peak
can be substantially suppressed.  Thus, for determination of $v$,
we take the most conservative approach and assume that the entire
observed peak could in principle be due to violation of BE
statistics. In this case, once again, $v \propto S$; uncertainties
in S and the calibration constant are as described above. This
yields a limit on the BE statistics violation parameter for
photons:
%--------------------------------------------------------------------
\begin{equation}
  v<1.2\cdot10^{-7} \; (90\%~c.l.).
\label{Eq9}
\end{equation}
%--------------------------------------------------------------------                     .              (9)

This represents the first result based on a new principle, which
in the ideal case gives a background-free signal arising from
violation of BE statistics for photons.  We believe that the limit
on $v$ can be decreased by several orders of magnitude, with
experiments based on this same principle but applying new
techniques (including the use of narrowband cw lasers and highly
efficient detection schemes). Such an experiment is now underway.

We thank C. Bowers, D. Brown, E. Commins, O. Greenberg, R.
Hilborn, L. Hunter, K. Jagannathan, S. Rochester, M. Rowe, M.
Suzuki, and M. Zolotorev for useful discussions; and R. Hilborn
also for the loan of major equipment.  This work was supported by
funds from Amherst College, Yale University, and NSF (grant
PHY-9733479).

%-----------------------------------------------------
\begin{figure}
\centerline{\psfig{figure=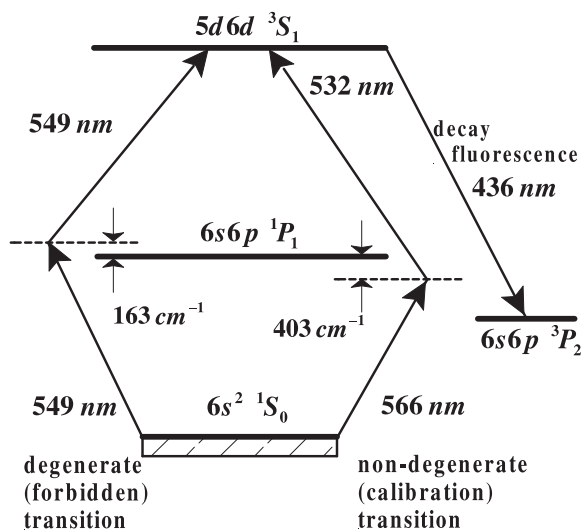,width=3.0 in}} \vspace{5
mm} \caption{ Excitation and detection schemes and relevant levels
in atomic barium.} \label{Fig1}
\end{figure}
%-----------------------------------------------------
%-----------------------------------------------------
\begin{figure}
\centerline{\psfig{figure=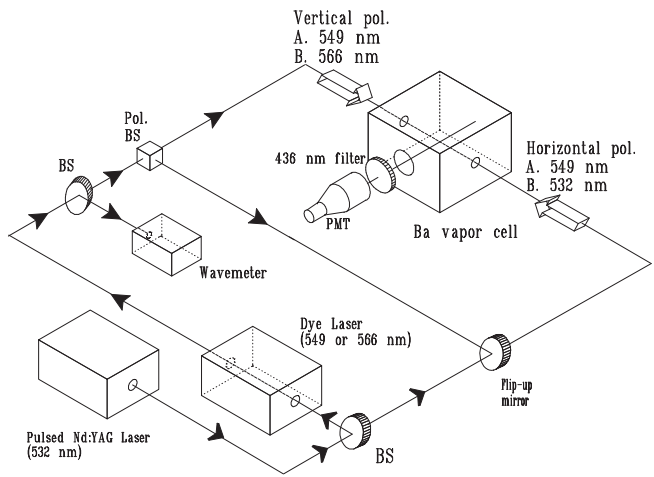,width=3.375 in}}\vspace{5
mm} \caption{ Schematic of the apparatus.  BS=beamsplitter.  For
the forbidden transition, the dye laser is tuned around $549~nm$,
and the flip-up mirror is in place as shown, so each of the
counterpropagating beams in the vapor cell originates from the dye
laser.  For the calibration transition, the dye laser is tuned to
$566~nm$ and the flip-up mirror is removed.  The beam entering the
vapor cell from the left then originates from the dye laser
($566~nm$), while the beam entering from the right originates from
the Nd:YAG laser ($532~nm$).} \label{Fig2}
\end{figure}
%-----------------------------------------------------
%-----------------------------------------------------
\begin{figure}
\centerline{\psfig{figure=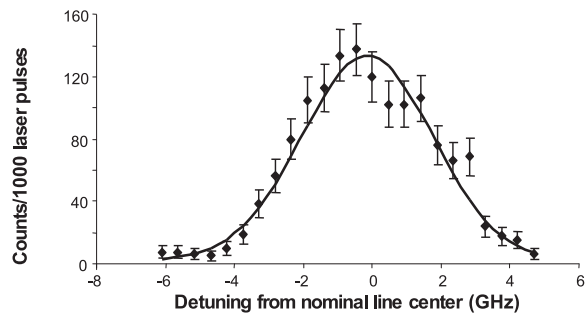,width=3.0 in}}\vspace{5 mm}
\caption{ Typical scan through the nondegenerate calibration
transition (points) and fit to determine peak height and linewidth
(solid line).  Taken with $230~\mu{}J/pulse$ at $566~nm$ and
$0.4~\mu{}J/pulse$ at $532~nm$.} \label{Fig3}
\end{figure}
%-----------------------------------------------------
%-----------------------------------------------------
\begin{figure}
\centerline{\psfig{figure=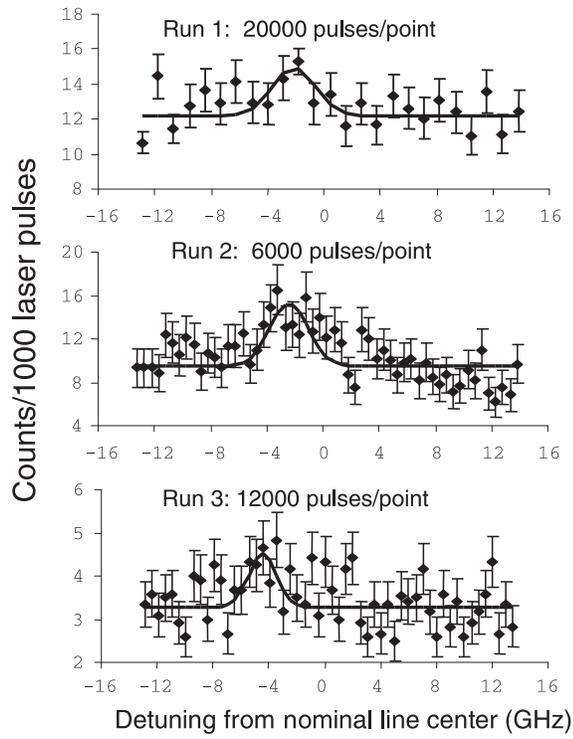,width=3.0 in}}\vspace{5 mm}
\caption{ Scans through the degenerate transition and best fits to
peak plus background.} \label{Fig4}
\end{figure}
%-----------------------------------------------------
\end{document}